\documentclass[12pt,a4paper]{article}
\pdfoutput=1
\usepackage{color}
\usepackage{jheppub}
\usepackage{mathtools,extarrows}

\usepackage{epstopdf}
\usepackage{graphicx}
\usepackage{epsfig}
\usepackage{dcolumn}  
\usepackage{bm}    
\usepackage{amssymb} 
\usepackage{amsmath,bm}
\usepackage{amsfonts}    
\usepackage{slashed}  
\usepackage{youngtab}
\usepackage[mathscr]{euscript}
\usepackage{epsfig}
\usepackage{verbatim}
\newdimen\tableauside\tableauside=1.0ex
\newdimen\tableaurule\tableaurule=0.4pt
\newdimen\tableaustep
\def\phantomhrule#1{\hbox{\vbox to0pt{\hrule height\tableaurule width#1\vss}}}
\def\phantomvrule#1{\vbox{\hbox to0pt{\vrule width\tableaurule height#1\hss}}}
\def\sqr{\vbox{%
		\phantomhrule\tableaustep
		\hbox{\phantomvrule\tableaustep\kern\tableaustep\phantomvrule\tableaustep}%
		\hbox{\vbox{\phantomhrule\tableauside}\kern-\tableaurule}}}
\def\squares#1{\hbox{\count0=#1\noindent\loop\sqr
		\advance\count0 by-1 \ifnum\count0>0\repeat}}
\def\tableau#1{\vcenter{\offinterlineskip
		\tableaustep=\tableauside\advance\tableaustep by-\tableaurule
		\kern\normallineskip\hbox
		{\kern\normallineskip\vbox
			{\gettableau#1 0 }%
			\kern\normallineskip\kern\tableaurule}%
		\kern\normallineskip\kern\tableaurule}}
\def\gettableau#1 {\ifnum#1=0\let\next=\null\else
	\squares{#1}\let\next=\gettableau\fi\next}

\newcommand{\be}{ \begin{equation}}
\newcommand{\ee}{\end{equation}}
\newcommand{\bea}[1]{\begin{eqnarray}\label{#1} }
\newcommand{\eea}{\end{eqnarray}}

\def\ZZZ{{\hskip-3pt\hbox{ Z\kern-1.6mm Z}}}
\def\zzz{{\hskip-3pt\hbox{ z\kern-1mm z}}}

\newcommand{\vp}{\varphi}

\def\bal#1\eal{\begin{align}#1\end{align}}

\def\one{{\hbox{ 1\kern-.8mm l}}}
\def\zero{{\hbox{ 0\kern-1.5mm 0}}}

\begin{document}

\title{Null hypersurface quantization, electromagnetic duality and asymptotic symmetries of Maxwell theory}

\author{Arpan Bhattacharyya${}^{a}$, Ling-Yan Hung${}^{a,b,c}$ and Yikun Jiang${}^{a,d}$}

\affiliation{$^a$ Department of Physics and Center for Field Theory and Particle Physics, Fudan University, \\
	\hspace*{0.3cm}220 Handan Road, 200433 Shanghai, P. R. China}

\affiliation{$^b$ State Key Laboratory of Surface Physics and Department of Physics, Fudan University,\\
	\hspace*{0.3cm}220 Handan Road, 200433 Shanghai, P. R. China}
	
\affiliation{$^c$ Collaborative Innovation Center of Advanced  Microstructures,
Nanjing University,\\
	\hspace*{0.3cm}Nanjing, 210093, P. R. China.}

\affiliation{$^d$ {Department of Applied Mathematics \& Theoretical Physics, University of Cambridge, \\
        \hspace*{0.3cm}Wilberforce Road, Cambridge CB3 0WA, United Kingdom}}

\emailAdd{
bhattacharyya.arpan@yahoo.com\quad elektron.janethung@gmail.com \quad phys.yk.jiang@gmail.com}

\abstract{In this paper we consider introducing careful regularization in the quantization of Maxwell theory in the asymptotic null infinity. This allows systematic discussions of the commutators in various boundary conditions, and application of Dirac brackets accordingly in a controlled manner. This method is most useful when we consider asymptotic charges that are not localized at the boundary $u\to \pm \infty$ like large gauge transformations. We show that our method reproduces the operator algebra in known cases, and it can be applied to other space-time symmetry charges such as the BMS transformations. We also obtain the asymptotic form of the U(1) charge following from the electromagnetic duality in an explicitly EM symmetric Schwarz-Sen type action. Using our regularization method, we demonstrate that the charge generates the expected transformation of a helicity operator. Our method promises applications in more generic theories. 
}



\maketitle

\makeatletter
\g@addto@macro\bfseries{\boldmath}
\makeatother

\section{Introduction}
In the pioneering work  \cite{Strominger1,Strominger2, He:2014laa,He1}, the quantization problem in the asymptotic boundary and the action of large gauge transformations were studied. Since then, much work has been done to study the relation between asymptotic symmetries and the infrared effects in various quantum field theories at null infinity ($\mathcal{I}^{\pm}$) \cite{ Lysov2, Campiglia3,Campiglia1,Campiglia2,Kapec1,Kapec3,Magnetic, Strominger3, Mirbabayi,Bousso,Campiglia5,Campiglia4,Dumitrescu,Nande,Hea,Gabai,Mitra} \footnote{The large gauge transformations generate charges on the radiative phase space \cite{Ashtekar,Ashtekar1} associated with these theories due to the presence of the massless particles. Much of the works mentioned above study the quantization of these charges.}. One crucial element in these discussions are related to the modification of the commutators for `zero modes' when extra boundary conditions like $F_{z \bar z}=0$ at $\mathcal{I}^{\pm}_{\pm}$ are imposed to recover a missing factor of $\frac{1}{2}$ in the transformation generated by the Noether charge. However, the inner workings of the method and its generalizations remain relatively mysterious,  partly because of the infinite volume of the asymptotic boundary (or the equivocal concept of the `zero mode' that occupy an infinitely small volume in the momentum space) that obscures the treatment. Notably,  it is not immediately obvious what is the appropriate treatment when we deal with

\begin{enumerate}
\item charges that are not localized at $u=\pm \infty$;

\item symmetry transform that are not compatible with the boundary conditions imposed, such as $F_{z \bar z}=0$ considered in  \cite{He:2014laa,He1}.
\end{enumerate}
In this paper, we would like to clarify the situation by introducing proper regularization and systematically obtain a new set of commutators in this controlled setting. When proper regularization is in place, it demonstrates with much more clarity the subtleties concerning quantization on a null hypersurface and the modification of the commutators. We will discuss these subtleties in detail, demonstrating how they can be dealt with systematically. We will also illustrate how Dirac brackets can be obtained as we impose further boundary conditions as constraints as in \cite{He:2014laa,He1}. These are the main results of the paper. Then we consider various  charges generating transformations corresponding to the asymptotic symmetries of the Maxwell theory, i.e by the  BMS \cite{ BBM, Sachs,Newman, Barnich} (or more generally conformal BMS  \cite{Haco}) generators and demonstrate that we can recover the expected commutation relations for the super-rotation and dilatation generators. However, additional subtleties arise while we deal with the super-translation and conformal BMS symmetries. What's more, it is noted \cite{He1} that the soft theorem implies the decoupling of a special linear combination of the two zero modes of different helicity from the S-matrix.  This motivates us to consider yet another symmetry of the Maxwell theory-- namely electromagnetic duality (EM duality)   which is deeply connected to helicity \cite{Zwanzig}  In the second half of the paper, we will use the methods developed in \cite{Schwarz, Pform} to construct an action equivalent to Maxwell action with manifest EM duality symmetry and quantize the theory at $\mathcal{I}^{\pm}$. One very important motivation is that the Noether charge involved does not generate the correct algebra using standard quantization. Extra constraints localized at the boundary points are imposed, and yet the charge itself is not localized at the boundary such that methods in \cite{He:2014laa,He1} are not readily applicable. 
This example therefore particularly illustrates the power of our method.

Before we end, let us emphasize that although this paper is focused on Maxwell theory, we believe this method promises many more applications in the quantization of generic theories at asymptotic infinity.

The organization of the paper is as follows. In section 2, we will revisit the quantization via `Schwinger quantization procedure' discussed in \cite{Frolov}. We will highlight the subtleties regarding boundary conditions in the procedure followed there, and explain why it is not suitable for the quantization problem on a null hypersurface with the boundary conditions that we are interested in. We will then show that our regularization procedure allows one to treat these issues explicitly, and obtain suitable commutators for the current situation.

In section 3, with our regularization in place, we will obtain commutators using the Schwinger method with different choices of boundary conditions. Extra boundary constraints can now be systematically and explicitly imposed via the Dirac procedure.  Then we apply our results to the problem studied in \cite{He:2014laa,He1}, and recover the algebra of the large gauge symmetry transformations there both with and without the extra boundary condition, showing that the method is consistent with existing results. 

Then we apply our methods to obtaining the correct commutation (sub)-algebra of asymptotic conformal BMS charges of the pure Maxwell theory in section 4. In section 5, we will describe the quantization of the Maxwell action that makes electromagnetic duality explicit. We will obtain the corresponding Noether charges and their commutators in the asymptotic region. 

A review of the Schwinger brackets that follow closely the discussions in \cite{Frolov} is relegated to the appendix. We devote extra emphasis on the subtleties of null hypersurface quantization and the importance  of keeping the contributions coming from the boundary generator $G_{b}$ (see equation (\ref{Gb}) ) that was previously set to zero as an extra constraint \cite{Frolov}. This is crucial particularly when we are interested in cases where the soft modes come into play. Then we conclude in the  section 6.

\section{Schwinger method and large gauge transformations}

In this section we will briefly discuss the Schwinger quantization procedure, and we will take large gauge transformations \cite{He1} as the first example. We will follow the procedure given in \cite{Frolov}, where  the Schwinger quantization procedure for  both  spacelike hypersurface and null hypersurface has been discussed in detail. Here, we only quote some important steps and show why they might not be suitable for problems related to soft modes. We mainly focus on the quantization on null hypersurface here and the details pertaining to spacelike hypersurface have been relegated to the appendix. Interested readers are referred to it for comparisons.

The current paper focuses on the quantization of the Maxwell theory at the asymptotic null infinity. The retarded coordinate which is suitable for describing physics at future null infinity ${\mathcal{I}^{+}}$ reads,
\be \label{eq2}
ds^2=-du^2-2 du dr+2 r^2 \gamma_{z\bar z} dzd\bar z,
\ee 
where, $u=t-r, \gamma_{z\bar z}=\frac{2}{(1+ z \bar z)^2}$, and $r$ is treated as `time' in the canonical quantization procedure, and `null infinity' means the region where $r\rightarrow \infty $.

The Maxwell action is
\be \label{action}
S=-\frac{1}{4}\int d^4x \sqrt{-g}\mathcal{F}^{\mu\nu}\mathcal{F}_{\mu\nu}.
\ee

We will work in the retarded radial gauge.
\begin{align} \label{gauge}
\begin{split}
&\mathcal{A}_{r}=0\\&
\mathcal{A}_{u}|_{\mathcal{I}^{+}}=\mathcal{O}(\frac{1}{r}),
\end{split}
\end{align}
and  the corresponding asymptotic behaviour of the gauge fields are given by \cite{He1}
\begin{align}   
\begin{split}\label{eq1003}
&\mathcal{A}_{z}(r,u,z,\bar z)=A_{z}(u,z,\bar z)+\sum_{n=1}^{\infty}\frac{A_{z}^{(n)}(u,z,\bar z)}{r^n},\\&
\mathcal{A}_{u}(r,u,z,\bar z)=\frac{1}{r} A_{u}(u,z,\bar z)+\sum_{n=1}^{\infty}\frac{A_{u}^{(n)}(u,z,\bar z)}{r^{n+1}}.
\end{split}
\end{align}
In the Schwinger quantization method, one first obtains a generator $G_{\Sigma}$ by varying the action. Then one requires that commutation with this generator recovers the transformation of the fields.  

\begin{align}
\begin{split} \label{eq3double}
G_{\Sigma}&=\int du dz d\bar z\Big(\delta A_{z}\partial_{u} A_{\bar z}+\delta A_{\bar z} \partial_{u}A_z\Big)-\frac{1}{2}\int dzd\bar z\Big(A_z \delta A_{\bar z}+A_{\bar z} \delta A_{z}\Big)\Big |^{\infty}_{-\infty}.
\end{split}
\end{align}
 We call the last term in (\ref{eq3double}) the boundary generator  
\be \label{Gb}
G_b=-\frac{1}{2}\int dzd\bar z\Big(A_z \delta A_{\bar z}+A_{\bar z} \delta A_{z}\Big)\Big |^{\infty}_{-\infty}.
\ee

It is important to note the  difference between quantization on a null hypersurface and a spacelike hypersurface. In the former case, the ``canonical momenta''  $F_{u \bar{z}}$ ($F_{u {z}}$)  are no longer independent of $A_{z}$ ($A_{\bar z} $), the gauge potentials themselves. In \cite{Frolov},  this problem was dealt with, by doing an integration by parts to separate the generator into a `bulk' term and a `boundary' term as in (\ref{eq3double}). Then taking only the fields $A_z$ and  $A_{\bar z} $ as independent degrees of freedom, and setting to zero variations of the fields at the boundary (i.e. setting $G_b=0$),  one demands, 
\be \label{eq100}
[A_{z},G_{\Sigma}]=\frac{i}{2} \delta {A_{z}},[A_{\bar z},G_{\Sigma}]=\frac{i}{2}\delta {A_{\bar z}}.
\ee
This implies that 
\begin{align} \label{eq1001}
\begin{split} 
  [A_{z}(u,z,\bar z), A_{\bar w} (u', w,\bar w)]=-\frac{i e^2}{4} \Theta(u-u')\delta^2 (z-w). \\
  [A_{z}(u,z,\bar z), A_{w}(u',w,\bar w)]= [A_{\bar z}(u,z,\bar z), A_{\bar w}(u',w,\bar w)]=0.
\end{split}
\end{align}
where $\Theta(u-u')$ is the sign function.

Subtleties arise when we consider cases where the boundary fields can also vary.  If $G_b$ do not vanish, then the commutators obtained in (\ref{eq1001}) gives
$$[A_{z},G_{b}]=[A_{\bar z},G_{b}]=0,$$  which means 
\begin{align}
\begin{split} \label{eq1000}
&\delta A_{z}(u=\infty,z,\bar z)+\delta A_{z}(u=-\infty, z, \bar z)=0\\
&\delta A_{\bar z}(u=\infty,z,\bar z)+\delta A_{\bar z}(u=-\infty, z, \bar z)=0
\end{split}
\end{align}
So the variations are not free, and we will have an extra constraint for our transformations. In the latter part of the original paper of \cite{Frolov,Ashtekar}, they imposed for simplicity that $\delta A$ vanishes at $u = \pm \infty$. This is possible had we treated large gauge transformations as gauge degrees of freedom as well. 

However, as the large gauge transformation should be treated as a real symmetry and that fields at $u=\pm \infty$ can still fluctuate (for example, the soft modes related to large gauge transformations), the commutators (\ref{eq1001}) is only applicable only if the constraints (\ref{eq1000}) are satisfied.

For example, for large gauge transformations,
\begin{align}
\begin{split} \label{largegauge}
&\delta_{\epsilon} A_{z}(u,z,\bar z)=\partial_{z}\epsilon(z,\bar z),\\&
\delta_{\epsilon}A_{\bar z}(u,\bar z,z)=\partial_{\bar z}\epsilon(z,\bar z).
\end{split}
\end{align}

We note that
\begin{align}
\begin{split} \label{eq99}
 &\delta A_{z}(u=\infty)+\delta A_{z}(u=-\infty)=2\partial_{z}\epsilon(z,\bar z) =0\\
 &\delta A_{\bar z}(u=\infty)+\delta A_{\bar z}(u=-\infty)=2\partial_{\bar z}\epsilon(z,\bar z) =0
\end{split}
\end{align}

The large gauge transformations are not compatible with the constraints (\ref{eq1000}). Consistent use of  (\ref{eq1001}) would require that the large gauge transformations be treated as a gauge degree of freedom. If we would like to include large gauge transformations as a genuine global symmetry in our theory, we need to modify the commutators and treat the boundary term consistently.

In \cite{He1}, a prescription has been provided to resolve the problem by introducing edge modes \cite{Mohd} and then imposing extra boundary conditions $F_{z \bar z}=0$ at $\mathcal{I}^{+}_{\pm}$ (both $r\rightarrow  \infty $ and $u \rightarrow \pm \infty$ ) to modify the commutators. But it is not clear how to generalize his methods beyond the special  boundary constraint $F_{z\bar z}=0$ at $\mathcal{I}^{+}_{\pm}$ considered there.  As we shall see, there are situations when $F_{z\bar z}\vert_{u\to\pm\infty}$ is not compatible with the symmetry of the problem, and that the expression for the Noether charge is not a $u$ total derivative (i.e when the charge is not localized at the boundary points and thus also getting contributions from non-zero modes). In these cases, the current treatment in the literature does not provide a clear and systematic procedure that singles out the changes to the commutators following from constraints localized at the end points of the Cauchy surface.

\section{A new regularization scheme for the self-consistent quantization of Maxwell theory on a null hypersurface}

In this section, we will show that there is a regularization scheme to include the contributions from the boundary modes automatically and make the quantization consistent even without imposing any extra constraints, and it is suitable for generic charges not localized at the boundary of null infinity.

Inspired by \cite{Zeromodes}, we first regularize our $u$ coordinate on a finite interval $[-\frac{T}{2}, \frac{T}{2}]$, and take $T \to \infty$ limit only at the end of the calculation. Then, we impose periodic boundary condition for the field strengths as 
\be\label{boundarycond}
F_{uz}(-\frac{T}{2})=F_{uz}(\frac{T}{2}), \qquad F_{u{\bar z}}(-\frac{T}{2})=F_{u{\bar z}}(\frac{T}{2}),
\ee which means that the radiated electric fields are zero in the far past and far future i.e.  $\int du\, F_u^{ z} F_{ uz} =0$. 
With this periodic boundary condition, we can expand our field strengths as 
\begin{align}
\begin{split}
 & F_{uz}(u,z,\bar z)=\sum_{m=-\infty}^{\infty}\alpha_{m}(z,\bar z) e^{\frac{i\,2\pi m u}{T}}\\
 & F_{u\bar z}(u,z,\bar z)=\sum_{m=-\infty}^{\infty}\bar \alpha_{m}(z,\bar z) e^{\frac{i\,2\pi m u}{T}}
\end{split}
\end{align}
from which we get the mode expansion of $A_{z}$ and $A_{\bar z}$ as

\begin{align}
\begin{split}\label{eq22}
&A_z(u,z,\bar z)=d_{0}(z,\bar z)+\alpha_{0}(z,\bar z) u+\sum_{m\neq 0}\frac{T}{i \, 2\pi\,m}\alpha_{m}(z,\bar z) e^{\frac{i\,2\pi m u}{T}},\\&
A_{\bar z}(u, z,\bar z) =\bar d_{0}(z,\bar z)+\bar \alpha_{0}(z,\bar z) u+\sum_{m\neq 0}\frac{T}{i \, 2\pi\,m}\bar \alpha_{m}(z,\bar z) e^{\frac{i\,2\pi m u}{T}}.
\end{split}
\end{align}

In the following, we will follow the Schwinger quantization procedure in the presence of this regularization. To obtain the commutators, we  substitute these regularized expressions into the charge $G_\Sigma$ given in (\ref{eq3double}). We note that since $A_z$ and $A_{\bar z}$ do not vanish at $u\to \pm T/2$, the separation between `boundary terms' and `bulk terms' become ambiguous. Therefore, one important departure from \cite{Frolov} is that we are obliged to keep these boundary terms (\ref{Gb}).
So, we end up having, 
  \begin{align}
  \begin{split}
 G_{\Sigma}=\frac{1}{2}\int dz d\bar z& \Big[T\,\bar \alpha_0 \delta d_0+T\, \alpha_{0}\delta \bar d_0 -\sum_{m\neq 0}\frac{i\,T^2}{m\pi}\Big(\bar \alpha_{m}\delta \alpha_{-m}+\alpha_{m}\delta \bar \alpha_{-m}\Big)\\&-\Big(T\, \bar d_0+\sum_{m\neq 0}\frac{i\, T^2 (-1)^m}{2\pi m}\bar \alpha_{m}\Big)\delta \alpha_0-\Big(T\, d_0+\sum_{m \neq 0}\frac{i\, T^2 (-1)^m}{2\pi m}  \alpha_m\Big)\delta \bar \alpha_0\\&+\sum_{m\neq 0}\frac{i\, T^2 (-1)^m}{2\pi m}\Big(\alpha_0\delta \bar \alpha_m+\bar \alpha_{0}\delta \alpha_m\Big)\Big].
   \end{split}
  \end{align}

Here, as these modes are independent, we demand, as in the Schwinger quantization procedure, 
\begin{align} \label{eq31}
[d_{0},G_{\Sigma}]=\frac{i}{2}\delta d_{0}, [\alpha_{0}, G_{\Sigma}]=\frac{i}{2}\delta \alpha_{0}, [\alpha_{m}, G_{\Sigma}]=\frac{i}{2}\delta \alpha_{m}
\begin{split}
\end{split}
\end{align}
and similarly for the $\bar d_{0}, \bar \alpha_{0}, \bar \alpha_{m}$. These relations are overdetermined, but pleasingly there is a set of consistent solutions, which gives,
\begin{align}
\begin{split} \label{eq32}
&[\bar d_{0}(z, \bar z) , \alpha_{0}(w,\bar  w )]=[d_{0}(z, \bar z) , \bar \alpha_{0}(w,\bar  w )]=\frac{i}{T}\delta^2 (z-w),\\&[\alpha_{m}(z,\bar z),\bar \alpha_{n}(w,\bar z)]=-\frac{m \pi}{T^2}\delta_{m+n,0}\delta^2(z-w),\\&[\bar d_{0}(z,\bar z),\alpha_{m}(w,\bar w)]=[ d_{0}(z,\bar z),\bar \alpha_{m}(w,\bar w)]=\frac{i}{2T}(-1)^m\delta^2(z-w).
\end{split}
\end{align}

We will use these commutators in the following sections to demonstrate how they work for different problems.
We note that open string quantization in the presence of a non-trivial $B$ field that leads to constraints at the end points were treated using a very similar procedure \cite{Chu:1998qz,Chu:1999gi}. The only subtlety in the current problem is the infrared limit $T\to \infty$ that has to be taken at the end.

\section{Quantization of large gauge transformations}
To begin with, we would like to ensure that our commutators recover the correct quantization conditions for large gauge transformations considered in \cite{He1}.
The large gauge transformations are given in (\ref{largegauge}).
The leading order Noether charge in the large $r$ expansion is 
\be
Q=\int du dz d\bar{z} \Big( \partial_{u} A_{\bar z} \partial_z \epsilon +\partial_{u} A_{z} \partial_{\bar z} \epsilon \Big).
\ee
In terms of our mode expansion, this becomes
  \be \label{soft}
  Q=\,T\,\int dz d\bar z \Big(\bar \alpha_0 \partial_{z}\epsilon+\alpha_0 \partial_{\bar z} \epsilon\Big).
  \ee
So we can see immediately using our commutators (\ref{eq32})
\begin{align}
\begin{split}
&[A_z(u,z,\bar z),Q]=i \partial_z \epsilon=i \delta A_z\\
&[A_{\bar z}(u,z,\bar z),Q]=i \partial_{\bar z} \epsilon=i \delta A_{\bar z}
\end{split}
\end{align}
which are the correct commutation relations that we would expect.
Here we note that the factor of $\frac{1}{2}$ problem as pointed out in \cite{He1} does not appear even if we are allowing $F_{z\bar z}$ to fluctuate freely at the boundary $u\to \pm T/2$. 
In the following, we would then further impose also the constraint  as in \cite{He1} 
\be \label{eq5}
F_{z\bar z}\Big|_{\pm \frac{T}{2}}=0.  
\ee
The constraint does not appear to be a necessary ingredient in recovering the correct commutators. We would like to study it in detail however to determine if our brackets would ultimately be consistent with  \cite{He1}. We would also like to demonstrate as an extra bonus, that Dirac brackets following from the constraints can be obtained in a systematic and transparent manner. 

Imposing these constraints in our system will modify our commutators defined in (\ref{eq32}). We will show how  this can be done explicitly using the Dirac procedure \cite{Henn}.  
Using our regularization, (\ref{eq5}) is equivalent to the following two constraints,
\begin{align}
\begin{split}\label{eq1002}
&\mathcal{\varphi}_{1}=\partial_{z}\bar \alpha_{0}-\partial_{\bar z} \alpha_0,\\&
\mathcal{\varphi}_{2}=\partial_{z}(\bar d_{0} +\sum_{m\neq 0}\frac{ (-1)^m T}{i\, 2\pi m}\bar \alpha_{m} )-\partial_{\bar z}(d_0+\sum_{m\neq 0}\frac{ (-1)^m T}{i\, 2\pi m} \alpha_{m}).
\end{split}
\end{align}
The modified commutators using Dirac procedure, for any two operators $F, G$ is obtained as follows,
\be
[F,G]^{D}= [F,G]-[F,\varphi_{\alpha}]C^{\alpha\beta}[\varphi_{\beta}, G],
\ee
where the matrix $C^{\alpha\beta}$ is the inverse matrix of 
\be
C_{\alpha\beta}=[\varphi_{\alpha}(z,\bar z),\varphi_{\beta}(w,\bar w)],
\ee
and $\alpha$ is the index for the constraints. Here, it takes value in $\{1,2\}$, and superscript $D$ denotes the modified commutators after imposing the constraints. 

Next we compute the $C_{\alpha\beta}.$ The non-vanishing components for the constraints (\ref{eq1002}) are 
\be
C_{12}=-C_{21}=[\varphi_1(z,\bar z),\varphi_2(w,\bar w)]=\frac{i}{T}(\partial_{z}\partial_{\bar w} +\partial_{\bar z}\partial_{w} )\delta^2(z-w).
\ee

So we will get, 

\begin{align}
\begin{split} \label{eq51}
&[\alpha_{0}(z,\bar z), \bar d_0(w,\bar w)]^{D}=-\frac{i}{2 T}\delta^2 (z-w)\\
&[\alpha_0(z,\bar z), d_{0}(w,\bar w)]^D=\frac{i}{4\pi T} \frac{1}{(z-w)^2} \,\\
&[\bar{\alpha_0}(z,\bar z),\bar{d_{0}}(w,\bar w)]^D=\frac{i}{4\pi T} \frac{1}{({\bar z}-{\bar w})^2} \,\\
&[d_0(z,\bar z),\bar \alpha_m(w,\bar w)]^D=[\bar{d_0}(z,\bar z), \alpha_m(w,\bar w)]^D=0\\
&[d_0(z,\bar z), \alpha_m(w,\bar w)]^D=-\frac{i}{4\pi T} \frac{(-1)^m}{(z-w)^2}\\
&[\bar{d_0}(z,\bar z), \bar{\alpha}_m(w,\bar w)]^D=-\frac{i}{4\pi T} \frac{(-1)^m}{(\bar z-\bar w)^2}
\end{split}
\end{align}
To compare with the charge given in \cite{He1},
\be \label{eq52}
Q=-2\int_{S^2} dz d\bar z \epsilon(z,\bar z) \,\,\partial_{z}\partial_{\bar z} (\phi_{+}-\phi_{-}),
\ee
we write $\phi_{+}$ and $\phi_{-}$ in terms of mode expansion,
\begin{align}
\begin{split}
&\partial_{z}\phi_{+}=d_0+\frac{T}{2}\alpha_0+\sum_{m\neq 0}\frac{T (-1)^m}{i\, 2\pi\,m}\alpha_{m},\\&\partial_{\bar z}\phi_{-}= \bar d_0-\frac{T}{2} \bar \alpha_0+\sum_{m \neq 0} \frac{T (-1)^m}{i\,2\pi\,m}\bar \alpha_{m}
\end{split}
\end{align}
Using (\ref{eq51}) it can be shown easily that,
\be \label{eq53}
[\phi_{+}(z,\bar z),\phi_{-}(w,\bar w)]=\frac{i}{4\pi}\ln|z-w|^2.
\ee
And from this, it automatically follows,
\be
[A_z(u,z,\bar z),Q]=i \partial_z \epsilon=i \delta A_z
\ee
which recovers the correct algebra again.

\subsection{Comments about soft photon theorems}

We end this section by briefly comment about soft photon theorems. We have quantized Maxwell theory with different boundary conditions, one that allow $F_{z\bar z}$ to fluctuate freely at $u\to \pm \infty$ and another, that is proposed in \cite{He1}, where $F_{z\bar z}\vert_{u\to \pm\infty} =0 $. While we found that using our regularization, both boundary conditions recover the correct large gauge transformation, it is not clear whether Weinberg's soft theorem -- which is demonstrated to be equivalent to the Ward identity following from the large gauge symmetry in \cite{He1} when $F_{z\bar z}\vert_{u\to \pm \infty} =0$ in \cite{He1} -- should carry through in different boundary conditions.

We note that by inspecting the statement of Weinberg's soft theorem, it is observed that a linear combination of soft photon decouples from the S-matrix. 
Quoting the appendix of \cite{He1}, the decoupled photon is given by
\be
a_{-}(\omega\to 0) - \frac{1}{2\pi} (1+ z \bar z) \int d^2 w \frac{1}{\bar z - \bar w} \partial_{\bar w}  \frac{a_+(\omega \to 0)}{1+ w\bar w}
\ee
and it behaves as if it is completely decoupled from the theory. In our regularized theory, this can be translated into the following form
\be \label{decouple}
 \bar\alpha_0  + \frac{1}{2\pi}\int d^2w \frac{1}{ \bar w - \bar z }  \partial_{\bar w} \alpha_0 ,
\ee
which is equivalent to the constraint $\vp_1$. In other words, Weinberg's soft theorem had already implied a choice of boundary condition at $u\to \pm\infty$, that sets $\vp_1$ to zero. 
Therefore, at least the constraint $\vp_1$ has to be imposed to be consistent with Weinberg's soft theorem, even without requiring  $\varphi_2$  constraint. So we only require $F_{z\bar z}$ to be periodic i.e $F_{z\bar z}(u=\frac{T}{2})=F_{z\bar z}(u=-\frac{T}{2})$ only not necessarilty strictly zero at $u\rightarrow \pm \frac{T}{2}.$ 
Imposing $\varphi_1$.

Without imposing any further constraints, however, one finds that the Ward identity again gives some other relations between the S matrix components with soft insertions and those without. The relation, however,  is different from Weinberg's soft theorem, as expected. It is interesting to notice however the ``two'' soft modes appearing in our Ward identity always appear together, so we cannot separate the two contributions and get ``soft theorems'' for each of them, due to the fact that the U(1) symmetry we break can only have at most one goldstone mode.

\section{Asymptotic symmetries of Maxwell action and the corresponding quantization}
It is well known that in 4 dimensions the Maxwell theory is conformally invariant \cite{Wald}. 

Given a four vector $\xi^{\rho}$, it generates space-time transformation for the field strengths as
\be
\delta \mathcal{F}_{\mu\nu}=\mathcal{L}_\xi F_{\mu\nu}= \xi^{\rho}\partial_{\rho}\mathcal{F}_{\mu\nu}+\partial_{\mu}\xi^{\rho} \mathcal{F}_{\rho\nu}-\partial_{\nu}\xi^{\rho} \mathcal{F}_{\rho\mu}.
\ee
where $\mathcal{L}_\xi$ is the Lie derivative with respect to $\xi$. Then the variation of the Maxwell action is given by
\be
\delta S=-\frac{1}{4} \int \sqrt{-g}d^4x \Big[\nabla_{\rho}(\mathcal{F}^{\mu\nu}\mathcal{F}_{\mu\nu}\xi^{\rho})-\frac{1}{2}[\mathcal{F}_{\mu\gamma}F_{\nu}{}^{\gamma}(\xi^{\rho}\partial_{\rho} g^{\mu\nu}+g^{\rho \mu}\partial_{\rho}\xi^{\nu}+g^{\rho\nu}\partial_{\rho} \xi^{\mu}-\frac{1}{2}g^{\mu\nu}\nabla_{\rho}\xi^{\rho})]\Big],
\ee
which vanishes when
\be \label{eq6}
\xi^{\rho}\partial_{\rho} g^{\mu\nu}+g^{\rho \mu}\partial_{\rho}\xi^{\nu}+g^{\rho\nu}\partial_{\rho} \xi^{\mu}-\frac{1}{2}g^{\mu\nu}\nabla_{\rho}\xi^{\rho}=0.
\ee
This is nothing but the conformal killing equations. \par 

However, this symmetry group  is enhanced at $\mathcal{I}^{\pm}.$  It has been shown in \cite{Haco} that the conformal symmetry group of the flat space time enhances to conformal BMS group at null infinity.  We next consider  the corresponding Noether charge for these asymptotic symmetries of the Maxwell theory and study their commutation relations. In the following we will  show that using our method we can recover the expected commutation relations for the dilatation and superrotation generators, while there are subtleties that remain for the supertranslation, BMS dilatation and BMS special conformal symmetries.  Simply speaking, for supertranslation, our regularization scheme introduces a linear $u$ term in the expansion of the gauge field that appears to break translation invariance in the $u$ direction, and the symmetry breaking has a remnant even in the limit $T \to \infty$. Although we do find a consistent way to get rid of the extra terms with a stronger boundary condition while still keeping the linear term, whether these are artifacts of our regularization scheme or a genuine symmetry breaking due to the appearance of the boundary modes should be studied in greater detail in the future. BMS dilatation and BMS special conformal transformations do not preserve the the gauge condition $\delta \mathcal{A}_{u}|_{\mathcal{I}^{\pm}}=\mathcal{O}(\frac{1}{r})$, even though they do preserve the boundary conditions for the field strength.  It is expected that they should be combined with an extra gauge transformation. We leave this also for future investigations. 

We will work with the boundary condition (\ref{boundarycond}) below.

\subsection{Dilatation}

For the conformal killing vector \cite{Haco},
\be
 \xi^{u}=u, \qquad \xi^r=r.
 \ee
the corresponding variation of the fields to leading order in the large $r$ expansion, is given by
 \be
 \delta A_z=u\partial_u A_z, \qquad \delta A_{\bar z}=u \partial_u A_{\bar z},
 \ee
 which scales as $r^0$.
 The corresponding charge at $\mathcal{I}^{\pm}$ is,
 \begin{align}
 \begin{split}
Q=\int du dz d \bar z \Big(F_{u\bar z} \delta A_{z}+F_{uz}\delta A_{\bar z}\Big)=2 \int du dzd\bar z\,   u\, (\partial_u A_z)(\partial_u A_{\bar z}).
  \end{split}
 \end{align}
 In terms of the mode expansion ,
 \be
 Q=\int dz d\bar z\Big[\sum_{n\neq 0}\frac{(-1)^nT^2}{i\, n\, \pi}\Big( \alpha_0\bar \alpha_n+\bar \alpha_0\alpha_n\Big)+\sum_{n \neq 0}\sum_{m \neq 0}\frac{(-1)^{m+n}T^2}{(m+n)\pi}\alpha_{m}\bar \alpha_n \Big].
 \ee
 Also one can check that it satisfy the correct quantization
\be
[A_z,Q]=i u\partial_u A_z=i \delta A_z
\ee
 using the commutators in  (\ref{eq32}).

\subsection{Superrotation}
We consider the asymptotic killing vector \cite{Barnich, Haco},
\be
\xi^u \equiv \frac{1}{2}u\, \psi, \qquad \xi^r=- \frac{1}{2}r\,\psi - \frac{1}{4}u\,D^2 \psi, \qquad \xi^A= Y^A - \frac{u}{2r}D^A \psi
\ee
where the superscript $A \in \{z, \bar z\}$; $D_A$ are the covariant derivatives with respect to the metric on the two-sphere; $Y^A$ are conformal Killing vectors on the 2-sphere, and $\psi=D_A Y^A$. 

Among these generators, the global part of the transformations are,
$Y^z=z^2, \,\, Y^{\bar z}=1; \,\, Y^z=1, \,\, Y^{\bar z}={\bar z}^2; \,\,Y^z=z,  \,\, Y^{\bar z}=-\bar z; \,\,\psi=0 $.

The corresponding Noether charge at $\mathcal{I}^{\pm}$ is,
\be
Q=\int du dz d \bar z \Big(F_{u\bar z} \delta A_{z}+F_{uz}\delta A_{\bar z}\Big),
\ee
where, up to leading order
\begin{align}
\begin{split}
&\delta A_z=\mathcal{L}_\xi A_{z}=\frac{1}{2} u \psi \partial_{u} A_z +Y^z \partial_{z} A_z+Y^{\bar z} \partial_{\bar z} A_z+A_z \partial_z Y^z,\\&
\delta A_{\bar z}=\mathcal{L}_\xi A_{\bar z}=\frac{1}{2} u \psi \partial_{u} A_{\bar z} +Y^z \partial_{z} A_{\bar z}+Y^{\bar z} \partial_{\bar z} A_{\bar z}+A_{\bar z} \partial_{\bar z} Y^{\bar z}. 
\end{split}
\end{align}
Using the mode expansion (\ref{eq22}) the charge looks like,
\begin{align}
\begin{split}
Q=\int dz d\bar z &\frac{1}{2} \psi \Big[\sum_{n\neq 0}\frac{(-1)^nT^2}{i\, n\, \pi}\Big( \alpha_0\bar \alpha_n+\bar \alpha_0\alpha_n\Big)+\sum_{n \neq 0}\sum_{m \neq 0}\frac{(-1)^{m+n}T^2}{(m+n)\pi}\alpha_{m}\bar \alpha_n \Big]\\
&+\Big[ \sum_{m\neq 0} \frac{ (-1)^m T^2}{i\,2 m\,\pi}  \bar  \alpha_m \Big(Y^z \partial_{z} +Y^{\bar z} \partial_{\bar z}+\partial_z Y^z\Big)\alpha_0\\
&-  \sum_{m\neq 0} \frac{T^2}{i\,2\pi\,m} \bar \alpha_m  \Big(Y^z \partial_{z} +Y^{\bar z} \partial_{\bar z}+\partial_z Y^z\Big)\alpha_{-m}+\bar \alpha_0 T  \Big(Y^z \partial_{z} +Y^{\bar z} \partial_{\bar z}+\partial_z Y^z\Big)d_0\\
&+ \sum_{m\neq 0} \frac{(-1)^m T^2}{i\, 2\pi\,m}\alpha_m \Big(Y^z \partial_{z} +Y^{\bar z} \partial_{\bar z}+\partial_{\bar z} Y^{\bar z}\Big)\bar \alpha_0\\
&-\sum_{m\neq 0}\frac{T^2}{i \, 2\pi\,m}\Big(Y^z \partial_{z} +Y^{\bar z} \partial_{\bar z}+\partial_{\bar z} Y^{\bar z} \Big) \bar  \alpha_{-m}  +\alpha_0 T\Big(Y^z \partial_{z} +Y^{\bar z} \partial_{\bar z}+\partial_{\bar z} Y^{\bar z}\Big) \bar d_0\Big].
\end{split}
\end{align}
Then using commutators defined in  (\ref{eq32}) and after some  algebraic manipulations, we can show that it also generates the correct transformation,
\be
[A_z,Q]=i \Big( \frac{1}{2} u \psi \partial_{u} A_z +Y^z \partial_{z} A_z+Y^{\bar z} \partial_{\bar z} A_z+A_z \partial_z Y^z \Big)=i \delta A_z.
\ee

 \subsection{Supertranslation}

 The supertranslation is generated by the following Killing vector \cite{Haco, Barnich, BBM,Sachs, Newman}
 \be
\xi^u \equiv f, \qquad \xi^A=- \frac{1}{r} D^A f
 \ee
where $f$ is any scalar spherical harmonic. And when $f$ is a constant, this is just the usual translation in the $u$ direction. The leading order term of the variation that will contribute to the charge is
\be
\delta A_z=\mathcal{L}_\xi A_{z}=f  \partial_{u} A_z, \delta A_{\bar z}=\mathcal{L}_\xi A_{z}=f  \partial_{u} A_z.
\ee
 The corresponding Noether charge at $\mathcal{I}^{\pm}$ is,
 \be
 Q=2 f \int du dz d\bar z(\partial_u A_z)(\partial_u A_{\bar z}) .
 \ee 
 In terms of the mode expansion,
 \be
 Q=2 f T \int dz d\bar z \Big(\alpha_0\bar \alpha_0+\sum_{m \neq 0} \alpha_m\bar \alpha_{-m}\Big).
 \ee
But one can show that the charge is not generating the expected transformation, i.e,
\be
[A_z,Q]\neq  i \delta A_{z}
\ee
Instead, we get
  \be {\label{eq999999999999}}
  [A_{z} (u,z,\bar z), Q]=i f \Big(\alpha_0(z,\bar z)+\sum_{m\neq 0}\alpha_{m}(z,\bar z)e^{\frac{i\,2\pi m u}{T}}\Big)+i f \Big(\alpha_{0}(z,\bar z)+ \sum_{m \neq 0}\,(-1)^m\alpha_{m}(z,\bar z) \Big).
  \ee
The first part corresponds to the expected transformation, however, we have an extra second part which is the boundary value of the first term.

What happened is that the regularization scheme we use introduces a linear $u$ term in the expansion of the gauge field that appears to break translation invariance in the $u$ direction.  We find that curiously, even in the limit $T \to \infty$ the symmetry breaking has a remnant when we study its operator algebra. Whether these are artifacts of our regularization scheme or a genuine symmetry breaking due to the appearance of the boundary modes should be studied in greater detail in the future. Especially, while the charge does recover the correct transformation for  $F_{uz}$ and $F_{u\bar z}$, it does not recover the expected transformation for $F_{z\bar z}$, where the linear term in $u$ remains. We note however, there are two possible remedies for the problem. We notice that the extra term in (\ref{eq999999999999}) is just the value of $F_{uz}\vert_{u\to \pm \infty}$. One solution isto impose extra boundary conditions with $F_{uz}\vert_{u\to \pm \infty}=F_{u\bar z}\vert_{u\to \pm \infty}=0$. This gets rid of the extra piece, recovering (super)translational invariance. There is another  solution, where the weaker boundary conditions $\partial_u F_{z\bar z}\vert_{u\to \pm \infty} = 0$ are imposed. This renders (\ref{eq999999999999}) a pure gauge, allowing us to modify our charge by adding to it the charge generating large gauge transformations to restore the correct transformation of the gauge potentials. We note that in both solutions, these extra boundary conditions do not contradict the condition $F_{z\bar z} = 0$ that would be a convenient boundary condition consistent with the soft theorem. Now that we have a systematic way of imposing constraints, we demonstrate a possible solution and obtain a new set of commutators consistent with $F_{uz}\vert_{u\to \pm \infty}=F_{u\bar z}\vert_{u\to \pm \infty}=0$ in the appendix.

\subsection{BMS dilatation and BMS special conformal transformation}
The BMS dilatation and special conformal transformations are tricky. The simple explanation is that they are mixing fields at different orders in $\frac{1}{r}$expansion, and they do not preserve the gauge condition we are considering for the gauge fields. To see that, 
we inspect the generators of the BMS dilatation and BMS special conformal transformations \cite{Haco}:
\begin{align}
\begin{split}
&\xi^u \equiv \frac{u^2}{2}, \qquad \xi^r= r(u+r)\\
&\xi^u \equiv \frac{u^2}{4} \zeta, \qquad \xi^r= - \left( \frac{u^2}{4} + \frac{r^2}{2} + \frac{u\,r}{2} \right)\zeta, \qquad \xi^A=- \frac{u}{2}\left(1 + \frac{u}{2r} \right)D^A\zeta 
\end{split}
\end{align}
where $\zeta$ are strictly conformal killing vectors of 2-sphere as compared to $\psi$ which are any killing vectors of 2-sphere. 

\subsubsection{Subtlety 1: commutators between different orders of the fields}
First of all, due to the $r^2$ term in the generator of the $\xi^r$ components, terms with different orders in the  $\frac{1}{r}$ expansion of the gauge fields mix together, for example, $\delta A_z$ is proportional to $A_z^{(1)}$ in equation (\ref{eq1003}).  
The operator algebra would require expansion in $\frac{1}{r}$ and  retain sub-leading order  term in $\mathcal{A}_z$ and also their corresponding commutation relations. Here, we will see that interestingly, further expanding the Schwinger generator shows that fields at different orders actually decouple (at least for order $\mathcal{O}(r^0)$ and $\mathcal{O}(r^{-1})$).

We look at the boundary term (\ref{eq3double})  more closely at this point.
\be
G_{\Sigma}=\frac{1}{2}\int du dz d\bar z r^2 \gamma_{z\bar z} \Big[\mathcal{A}_{z}\delta \mathcal{F}^{r z}+\mathcal{A}_{\bar z} \delta \mathcal{F}^{r\bar z}+\mathcal{A}_{u}\delta \mathcal{F}^{r u}-\delta \mathcal{A}_{z} \mathcal{F}^{r z}-\delta \mathcal{A}_{\bar z} \mathcal{F}^{r\bar z}-\delta \mathcal{A}_{u} \mathcal{F}^{r u}\Big].
\ee
Also we note , $\mathcal{F}^{r z}=\frac{1}{r^2 \gamma_{z\bar z}}(\mathcal{F}_{r \bar  z}-\mathcal{F}_{u \bar z}), \mathcal{F}^{ru}=\mathcal{F}_{ur}$. Using these facts one can easily check that the $\mathcal{O}(r^0)$ term is equal to  (\ref{eq3double}). Now the $\mathcal{O}({r}^{-1})$ contribution is,
\begin{align}
\begin{split} {\label{eq20000}}
G^{(1)}_{\Sigma}=\frac{1}{2 r}\int du dz d\bar z &\Big(-A^{(1)}_z \partial_u \delta A_{\bar z}-A_z \partial_u \delta A^{(1)}_{\bar z}-A^{(1)}_{\bar z}\partial_u\delta A_z-A_{\bar z}\partial_u \delta A^{(1)}_{z}\\&+\partial_u A^{(1)}_{z}\delta A_{\bar z} +\partial_u A_z \delta A^{(1)}_{\bar z}+\partial_u A^{(1)}_{\bar z}\delta A_z+\partial_{u} A_{\bar z} \delta A^{(1)}_{z}\\
&-\partial_z A_u\delta A_{\bar z}-\partial_{\bar z} A_u \delta A_z+A_{\bar z} \partial_{z}\delta A_u +A_{z}\partial_{\bar z}\delta A_u \Big),
\end{split}
\end{align}
where, we have used the expansion  (\ref{eq1003}).  Note that the  contribution coming from $\mathcal{A}_{u}\delta \mathcal{F}^{r u}$ and $\delta \mathcal{A}_{u}\mathcal{F}^{r u}$ cancel among themselves.

One of the equations of motion at leading order in $r$ reads,
\be
\gamma_{z\bar z}\partial_{u} A_{u}=\partial_{u}(\partial_{z}A_{\bar z}+\partial_{\bar z} A_{z}).
\ee
Which means 
\be
\gamma_{z\bar z}  A_{u}=\partial_{z}A_{\bar z}+\partial_{\bar z} A_{z} +h(z, \bar{z}) .
\ee
where $h(z,\bar z)$ is the integration constant of $u$. As we will see in the next section, this condition is relating $A_u$ directly to `the dual gauge potential $C$'. This is expected as this equation of motion is relating the electric and magnetic fields. Like in \cite{Magnetic}, we set all integration constants to be zero for simplicity, thus making $$h(z,\bar z)=0, \, A_u=\frac{\partial_{z}A_{\bar z}+\partial_{\bar z} A_{z}}{\gamma_{z\bar z}}.$$

Using this, the third line of (\ref{eq20000}) vanishes, as it will become total derivatives of $z$ and $\bar z$.

In the spirit of Schwinger quantization procedure, we require the following condition valid for all orders of $r,$
\be \label{Gtrans}
[\mathcal{A}_{w}, G_{\Sigma}]=\frac{i}{2}\delta \mathcal{A}_{w}.
\ee
At $\mathcal{O}(\frac{1}{r})$ we will have,
\be \label{Gtrans1}
[A_{w}, G^{(1)}_{\Sigma}]+[\frac{A^{(1)}_w}{r}, G^{(0)}_{\Sigma}]=\frac{i}{2}\delta \frac{A^{(1)}_{w}}{r}.
\ee

We note that the commutators among the leading order terms in (\ref{Gtrans}) already recovers the correct transformation of $\delta A^{(1)}_{z, \bar{z}}$. Therefore one would require that the brackets that follow from (\ref{Gtrans1}) do not alter the result of $\delta {A^{(1)}_{z,\bar{z}}}$.   One natural solution is that different orders decouple, i.e
\be \label{virtue}
[A_w^{(1)}, A_z]=0, [A^{(1)}_w, A_{\bar z}]=0.
\ee 

This procedure can be generalized to arbitrary  orders in $\mathcal{O}(\frac{1}{r})$ terms. 
Solving for the complete set of brackets is beyond the scope of the current paper, and we end here only with the roadmap of how it can be obtained at arbitrary order. 
We caution that there could potentially exist other solutions, although the computation strongly suggests that this is the correct solution. 

\subsubsection{Subtlety 2: An extra gauge transformation}

There is one more subtle issue, namely, that the two transformations do not preserve the gauge condition for $\mathcal{A}_u$, i.e, the leading order $1/r$ asymptotic behaviour of  $\mathcal{A}_u$.

For the BMS dilatation, we have
\be
\delta{\mathcal{A}_u}= - A_u .
\ee
For the BMS special conformal transformation,
\be
\delta{\mathcal{A}_u}=- A_u- \frac{1}{2}A_z D^z \xi -\frac{1}{2}A_{\bar z} D^{\bar z} \xi .
\ee 

The transformations are both of order $\mathcal{O}(r^{0})$, violating $\mathcal{A}_{u}|_{\mathcal{I}^{+}}=\mathcal{O}(\frac{1}{r})$. We notice that it is possible to recover the gauge condition using a further gauge transformation. This is possible because the leading term in the asymptotic expansion of the  gauge invariant field strengths are invariant under the transformations generated by the Lie derivatives of the corresponding asymptotic conformal killing vectors, unlike the gauge fields. 
We leave a complete discussion of the problem for future research.

\section{Electromagnetic duality and its quantization}

In this section we will consider yet another symmetry of the Maxwell theory - ` Electromagnetic  duality' (EM duality). 

In the pure Maxwell theory, the Maxwell equations clearly remain invariant when we exchange the electric and magnetic fields  $\vec E \rightarrow \vec B$ and $\vec B \rightarrow -\vec E$. This is only a discrete subgroup of the complete electric-magnetic duality of Maxwell equations. It is well known that in fact the equations of motions are invariant under a continuous SO(2) rotation $\delta \vec E=\theta \vec B, \delta \vec B =-\theta \vec E$. However, the Maxwell action itself, which is simply $\int (\vec E^2-\vec B^2)$, does not remain invariant under the transformation. 
In 1968, Zwanziger\cite{Zwanzig} first introduced a `dual potential' method to make the symmetry explicit at the level of the action, but at the price of giving up locality in the symmetry transformation. Interestingly, the author noticed that the conserved charge generating the symmetry after quantization becomes the total helicity of photons, and it is the number operator for right handed photons minus the left handed one(which he called chirality in his original paper, but for massless particle, we know that helicity is equivalent to chirality, and we take the latter name in this paper).  Later in \cite{Schwarz} it has been observed that there is a way using the techniques introduced in  \cite{Pform} to construct a local duality invariant action by introducing an auxiliary field such that the theory can be reduced to the usual Maxwell theory by eliminating the auxiliary field using the non-dynamical equations of motion, which are nothing but the duality condition. This method provides a perfect way to  manifest the duality invariance explicitly while losing manifest Lorentz covariance.

As noticed in \cite{He1}, the soft theorem implies that a special linear combination of two zero modes of different helicities decouple from the S-matrix. This motivates us to consider the EM duality symmetry, which is deeply connected to helicity, also at null infinity. Another motivation of considering this symmetry is that this is a typical symmetry for which the corresponding charge density does not localize at the boundary, and that the symmetry transformation is not compatible with the boundary condition  $F_{z \bar z}=0$ at $\mathcal{I}^{+}_{\pm}$. Not surprisingly, if we naively apply equation  (\ref{eq1001}), the commutation rules are incorrect due to the problematic treatment of soft-modes. 
We will deal with this issue using the regularization methods developed in the previous sections. 

\subsection{Duality invariant action for null coordinate}
To proceed, we obtain the Noether charge of this symmetry by following the approach of \cite{Schwarz, Pform}.
As the action in \cite{Schwarz} has lost Lorentz covariance, we will first derive a duality invariant action which is adapted to the coordinate system defined in (\ref{eq2}). We will first build the action in the gauge $\mathcal{A}_r=\mathcal{C}_r=0$ (where $\mathcal{C}_\mu$ is the dual field), and then focus on retarded radial gauge where the corresponding asymptotic behaviour of the gauge fields are given by (\ref{gauge} - \ref{eq1003}). 

Starting with the naive action with two copies of the gauge fields,
\be \label{eq4}
S_{em}=-\frac{1}{8}\int d^{4} x r^2\gamma_{z\bar z}\Big[ \mathcal{F}_{\mu\nu}\mathcal{F}^{\mu\nu}+\mathcal{G}_{\mu\nu}\mathcal{G}^{\mu\nu}\Big].
\ee

where $\mathcal{F}_{\mu\nu}$ is the field strength corresponding to the gauge field $\mathcal{A}_{\mu}.$ $\mathcal{G}_{\mu\nu}$ is the field strength corresponding to the dual field $\mathcal{C}_{\mu}$. 

Then we would like to impose the following duality condition,
\be
\label{eq41}
\frac{1}{2}\hat \epsilon_{\mu\nu\rho\sigma}\mathcal{F}^{\rho\sigma}=\mathcal{G}_{\mu\nu}.
\ee
where $\hat \epsilon_{\mu\nu\rho\sigma}$ is the Levi-Civita tensor with each of its non vanishing components taking values either $i$ and $-i$ in the coordinate system we have adopted\footnote{In our convention $\epsilon_{urz\bar z}=i$}.
With this duality condition imposed, (\ref{eq41}) gives us back the original Maxwell action in terms of the only gauge fields $\mathcal{A}_{\mu}.$ 

Explicitly, they are
\begin{align}
\begin{split} \label{eq200000}
&\frac{1}{2}\mathcal{F}_{ru}r^2\gamma_{z\bar z}=-\frac{i}{2}\mathcal{G}_{z\bar z},\qquad \frac{1}{2}\Big[\mathcal{F}_{u\bar z}-\mathcal{F}_{r\bar z}\Big]\,=\frac{i}{2}\mathcal{G}_{u\bar z},\\
& \frac{1}{2}\Big[\mathcal{F}_{uz}-\mathcal{F}_{r z}\Big]=-\frac{i}{2}\mathcal{G}_{uz}, \qquad\frac{i}{2}\mathcal{F}_{z\bar z}=\frac{1}{2}\mathcal{G}_{ru}r^2\gamma_{z\bar z}\,,\\
&-\frac{i}{2}\mathcal{F}_{u\bar z}=\frac{1}{2}\Big[\mathcal{G}_{u\bar z}-\mathcal{G}_{r\bar z}\Big], \qquad \frac{i}{2}\mathcal{F}_{uz}=\frac{1}{2}\Big[\mathcal{G}_{u z}-\mathcal{G}_{r z}\Big].
\end{split}
\end{align}

However, to avoid introducing extra degrees of freedom into our theory, we must impose the duality conditions as non-dynamical constraints arising naturally for the action. 
Our following treatment is inspired by \cite{Pform} and adopted and modified according to our current problem. 
To that end, we first write down the hamiltonian corresponding to (\ref{eq4}).
\be \label{eq42}
\mathcal{H}=\frac{(\pi_{A}^{u})^2}{r^2 \gamma_{z\bar z}}-\frac{1}{4}\frac{(\mathcal{F}_{z\bar z})^2}{r^2 \gamma_{z\bar z}} -\frac{1}{2}\mathcal{F}_{uz}\mathcal{F}_{u\bar z}+\pi^{\bar z}_{A}\mathcal{F}_{u\bar z}+\pi^{z}_{A}\mathcal{F}_{uz}-2\pi^{\bar z}_{A}\pi^{z}_{A}+dual.
\ee
By ``dual'' we mean that $\mathcal{F}_{\mu\nu}$ and $\pi^{\mu}_{A}$ are replaced by $\mathcal{G}_{\mu\nu}$ and $\pi^{\mu}_{C}.$ The conjugate momenta are defined as,
\begin{align}
\begin{split} \label{eq43}
\pi^{u}_{A}&=\frac{1}{2}\mathcal{F}_{ru}r^2\gamma_{z\bar z},\qquad \pi^{z}_{A}=\frac{1}{2}\Big[\mathcal{F}_{u\bar z}-\mathcal{F}_{r\bar z}\Big]\,,\\
\pi^{\bar z}_{A}&=\frac{1}{2}\Big[\mathcal{F}_{uz}-\mathcal{F}_{r z}\Big], \qquad\pi^{u}_{C}=\frac{1}{2} \mathcal{G}_{ru}r^2\gamma_{z\bar z}\,,\\
\pi^{z}_{C}&=\frac{1}{2}\Big[\mathcal{G}_{u\bar z}-\mathcal{G}_{r\bar z}\Big], \qquad \pi^{\bar z}_{C}=\frac{1}{2}\Big[\mathcal{G}_{u z}-\mathcal{G}_{r z}\Big].
\end{split}
\end{align}
Next we impose (\ref{eq41}) 
on (\ref{eq43}) and rewrite them as
\begin{align}
\begin{split} \label{eq44}
\pi^{u}_{C}&=\frac{i}{2}\mathcal{F}_{z\bar z},\qquad \pi^{u}_{A}=-\frac{i}{2}\mathcal{G}_{z\bar z}, \qquad \pi^{\bar z}_{A}=-\frac{i}{2}\mathcal{G}_{uz}\,,\\
\pi^{z}_{A}&=\frac{i}{2}\mathcal{G}_{u\bar z}, \qquad \pi^{\bar z}_{C}=\frac{i}{2}\mathcal{F}_{uz}, \qquad \pi^{z}_{C}=-\frac{i}{2}\mathcal{F}_{u\bar z}.
\end{split}
\end{align}
Then we take this modified Hamiltonian to recover an action again by using (\ref{eq44}) to get
\begin{align}
\begin{split}\label{eq1004}
S=\int d^{4}x&[\pi^{u}_{A}\partial_{r} \mathcal{A}_{u}+\pi^{z}_{A}\partial_{r}\mathcal{A}_{z}+\pi^{\bar z}_{A}\partial_{r}\mathcal{A}_{\bar z}+\\
    &\pi^{u}_{C}\partial_{r} \mathcal{C}_{u}+\pi^{z}_{C}\partial_{r}\mathcal{C}_{z}+\pi^{\bar z}_{C}\partial_{r}\mathcal{C}_{\bar z}-\mathcal{H}]\\
=\int d^{4}x \Big[& -\frac{i}{2}\mathcal{G}_{z\bar z}\mathcal{F}_{ru}+\frac{i}{2}\mathcal{G}_{u\bar z}\mathcal{F}_{rz}-\frac{i}{2}\mathcal{G}_{uz}\mathcal{F}_{r\bar z}+
\\&\frac{i}{2}\mathcal{F}_{z\bar z} \mathcal{G}_{ru}-\frac{i}{2}\mathcal{F}_{u\bar z} \mathcal{G}_{r z}+\frac{i}{2}\mathcal{F}_{uz}\mathcal{G}_{r\bar z}+\\&\frac{1}{2}\frac{(\mathcal{F}_{z\bar z})^2}{r^2 \gamma_{z\bar z}}+\frac{1}{2}\frac{(\mathcal{G}_{z\bar z})^2}{r^2 \gamma_{z\bar z}}+\mathcal{F}_{uz}\mathcal{F}_{u\bar z}+\\&\mathcal{G}_{uz}\mathcal{G}_{u\bar z}-i \mathcal{F}_{uz}\mathcal{G}_{u \bar z}+i \mathcal{F}_{u\bar z}\mathcal{G}_{uz}\Big].
\end{split}
\end{align}
Note that for this action, we still have individual gauge transformations $\delta \mathcal{A}_i=\partial_i \Lambda^{(A)}$ and $\delta \mathcal{C}_i=\partial_i \Lambda^{(C)}$ for the two gauge fields.

Next we find  that the variation of this action with respect to $\mathcal{C}_{u}$ gives the following equations of motion, 
\be
\partial_{z}\Big[\mathcal{F}_{u\bar z}-\mathcal{F}_{r \bar z}-i\mathcal{G}_{u \bar z}\Big]+\partial_{\bar z}\Big[\mathcal{F}_{r z}-\mathcal{F}_{u z}-i\mathcal{G}_{u z}\Big]=0.
\ee
Similarly for $\mathcal{C}_{z}$ and $\mathcal{C}_{\bar z}$ we get,

\be
\partial_{u}\Big[\mathcal{F}_{r\bar z}-\mathcal{F}_{u\bar z}+i\,\mathcal{G}_{u \bar z}\Big]+\partial_{\bar z}\Big[\mathcal{F}_{ru}+\frac{i}{r^2\gamma_{z\bar z}} \mathcal{G}_{z\bar z}\Big]=0.
\ee 
and,

\be
\partial_{u}\Big[\mathcal{F}_{rz}-\mathcal{F}_{uz}-i\,\mathcal{G}_{uz}\Big]-\partial_{z}\Big[\mathcal{F}_{ru}+\frac{i}{r^2\gamma_{z\bar z}} \mathcal{G}_{z\bar z}\Big]=0.
\ee 
By taking suitable linear combinations, we get,
\begin{align}
\begin{split}
&\partial_{u}\partial_{\bar z}\Big[\mathcal{F}_{r z}-\mathcal{F}_{u z}-i\mathcal{G}_{u z}\Big]=0,\\&
\partial_{z}\partial_{\bar z} \Big[\mathcal{F}_{ru}+\frac{i}{r^2\gamma_{z\bar z}} \mathcal{G}_{z\bar z}\Big]=0\\&
\partial_{u}\partial_{ z}\Big[\mathcal{F}_{u\bar z}-\mathcal{F}_{r \bar z}-i\mathcal{G}_{u \bar z}\Big]=0.
\end{split}
\end{align}

Just as in \cite{Schwarz}, these equations of motion are exactly the duality conditions (\ref{eq41}) we want to impose when $\Lambda^{(2)}$ is zero in the ``temporal'' gauge $\mathcal{A}_r=\mathcal{C}_r=0$. As now it is clear that the equations of motion for the $\mathcal{C}$ fields do not involve any $r$ derivative, (or in other words, they become constraints for our new Lagrangian \cite{Henn}), we can treat them as auxiliary fields and eliminate them by substituting non-dynamical equations of motion. Then  finally we can recover the usual Maxwell action (\ref{action}) in the gauge $\mathcal{A}_r=0$.

It is interesting to note that, from the above  equations of motion at leading order in $r$  we have,
\begin{align}
\begin{split} 
&\partial_u (C_z-i A_z)=0\\
&\partial_u (C_{\bar z}+i A_{\bar z})=0
\end{split}
\end{align}
So we have
\begin{align}
\begin{split} \label{eq46}
&C_{z}= i A_{z} +f (z,\bar z),\\&
C_{\bar z}= -i A_{\bar z} +\bar f (z,\bar z),
\end{split}
\end{align}
where $f(z, \bar z)$ and $\bar f (z,\bar z)$ are arbitrary functions of $z$ and $\bar z$. And if we take them to be zero as in \cite{Magnetic}, these equations have a very neat form, which is also what we did in the previous section. However, we will keep them in the next section to show that the quantization is consistent even when these fields are there.

\subsection{Electromagnetic duality and quantization at null infinity}

Now we can show that the following transformation,
\begin{align}
\begin{split}  \label{eq47}
&\delta A_{z}=\theta\, C_{z},\qquad \delta A_{\bar z}=\theta\, C_{\bar z},\\
&\delta C_{z}=-\theta\, A_{z}, \qquad \delta C_{\bar z}=-\theta\, A_{\bar z}
\end{split}
\end{align}
where $\theta$ is a constant, keeps the action (\ref{eq1004}) invariant at $\mathcal{I}^{\pm}$.  This is a realization of EM duality via a local symmetry transformations. 
 We arrive at the following Noether charge corresponding to the transformations (\ref{eq47}) after using equations of motion (\ref{eq46})
\be \label{eq1005}
Q=i\,\theta \int du \int dz d\bar z \Big[(\partial_{u} A_{\bar z})A_{z}-(\partial_{u}A_z) A_{\bar z}\Big]+\frac{\theta}{2}\int dz d \bar z \Big[ f A_{\bar z}+ \bar f A_{z}\Big]\Big|_{-\infty}^{\infty}. 
\ee

It's interesting to notice that if we use the commutators of (\ref{eq1001}), we will get
\be
 [A_z(u,z,\bar z),Q]=-\theta A_{z}(u,z,\bar z)+\frac{i\,\theta}{4} f(z,\bar z)+ \frac{\theta}{4}\Big( A_{z}(\infty,z,\bar z)+A_{z}(-\infty,z,\bar z)\Big).
\ee
rather than the expected
\be
[A_z,Q]=i \delta A_{z}=\theta\, C_{z}=-\theta\, A_z+i \, \theta\,f.
\ee
The problematic parts are all related to the boundary modes, i.e, those soft modes in momentum space.

On the other hand, if we impose the boundary condition $F_{z \bar z}=0$ at $\mathcal{I}^{+}_{\pm}$, all the degrees of freedom are removed.
To see this, we take $f=\bar f=0$ for simplicity, then in order to have $\delta F_{z \bar z}=0$ such that we still preserve the boundary condition after the transformation, we must have $F_{z \bar z}$ and $G_{z \bar z}$ both zero at $\mathcal{I}^{+}_{\pm}$, equivalently, it means that we have four constraints,

\begin{align}
\begin{split}\label{eq1010}
&\mathcal{\varphi}_{1}=\partial_{z}\bar \alpha_{0}\\
&\mathcal{\varphi}_{2}=\partial_{\bar z} \alpha_0,\\
&\mathcal{\varphi}_{3}=\partial_{z}(\bar d_{0} +\sum_{m\neq 0}\frac{ (-1)^m T}{i\, 2\pi m}\bar \alpha_{m} )\\
&\mathcal{\varphi}_{4}=\partial_{\bar z}(d_0+\sum_{m\neq 0}\frac{ (-1)^m T}{i\, 2\pi m} \alpha_{m}).
\end{split}
\end{align}

Using the Dirac procedure, we can show that all commutators become zero, thereby removing all the remaining degrees of freedom. We will thus relax this constraint in the following.

Now substituting into (\ref{eq1005}) the mode expansion, we have
\begin{align}
 \begin{split} \label{eq48}
 Q=&i\, \theta\int d^2z \,\, \Big[\bar \alpha_{0}\, d_{0}\, T-\alpha_{0}\,\bar  d_{0}\, T+ \sum_{n\neq 0}\frac{T^2}{i  \pi n}\alpha_{n}\bar  \alpha_{-n} -\sum_{m\neq 0}\frac{i (-1)^m T^2}{2 \pi  m}(\alpha_{0}  \bar  \alpha_{m} -\bar  \alpha_{0} \alpha_{m})\Big]\\&+\frac{\theta}{2}\int dz d\bar z\Big[T(f\, \bar \alpha_0+ \bar f\,\alpha_0)\Big].
  \end{split}
 \end{align}
We impose only the boundary condition (\ref{boundarycond}). One  can easily check that (\ref{boundarycond}) is consistent with the transformation (\ref{eq47}). Then using the commutators defined in (\ref{eq32}) we can easily check that,
\be \label{QAtrans1}
[A_z,Q]=-\theta\, A_z+i \, \theta\,f=i \delta A_z,
\ee
and similarly 
\be \label{QAtrans2}
[A_{\bar z},Q]=\theta\, A_{\bar z} + i \, \theta\,\bar f= i \delta A_{\bar z}.
\ee
So we have correctly quantized our charge. \\

Now we would like to inspect the ``EM duality charge'' of individual modes in the expansion. We notice that
they satisfy 
\be
[\alpha_n, Q] = - \theta \alpha_n, \qquad [\bar \alpha_n, Q]  = +\theta \bar \alpha_n;
\ee
which follows directly from (\ref{QAtrans1},  \ref{QAtrans2}). This fits very well with the known helicity of these modes. Recall that 
\be
\alpha_{n<0} = \bar{\alpha}^\dag_{n>0}, \qquad \bar\alpha_{n<0} = \alpha^\dag_{n>0},
\ee
we then identify $\alpha_{n>0}, \alpha^\dag_{n>0}$ to be the creation and annihilation operators respectively for positive helicity modes, and $\bar\alpha_{n>0}, \bar\alpha^\dag_{n>0}$ to be negative helicity modes. 

Finally, we can identify the decoupled soft photons in our formalism. Comparing with \cite{He1}, the decoupled mode is given by the linear combination (\ref{decouple}),
or equivalently as we discussed previously,
\be
\partial \bar \alpha_0 - \bar\partial \alpha_0,
\ee
which is indeed a linear combination of the zero modes with opposite helicity.

This gives support to the proposal that the U(1) charge following from electro-magnetic duality does behave in the expected way as helicity.

\section{Summary and discussions}

The main result of this paper is to develop a regularization method that allows one to study the commutators in a controlled and systematic way in the asymptotic null infinity, where quantization on a null hypersurface is particularly subtle. We demonstrate the power of our method in recovering known algebra of the large gauge transformations with different boundary conditions, and demonstrate that extra constraints can be imposed using Dirac brackets in a very transparent manner.  Then we push our method to more general symmetry charges, including various asymptotic space-time symmetries notably the BMS symmetry where the symmetry charge is no longer localized at the boundary points. We demonstrate that apart from the main subtle situation,  namely when the symmetry transformation is not immediately preserved by the boundary conditions, such as  the case of translation symmetry in $u$ which is broken by the linear $u$ term in the regularized mode expansion, our method reproduces the expected algebra. It appears that this can be fixed by imposing some mild constraints at $u\to \pm\infty$ which we demonstrate in detail in the appendix.  Finally, we highlight the application of our method in quantizing the Maxwell theory where electromagnetic duality is made explicit in the Schwarz - Sen type action.  This allows us to derive a conserved charge for electromagnetic duality, which, for a long time, is believed to be related to helicity. Here, we demonstrate that using our regularization method, we recover the expected commutators at asymptotic infinity, and that the quantum numbers with respect to the duality charge does coincide with the known helicity of the modes. The decoupled soft mode also acquires a very simple form in our regularized expansion. 

Our paper illustrates a very general procedure, that does not depend on the precise theory at hand. This promises applications in many other situations, such as in the study of operator algebra for gravitons in the asymptotic infinity, and the search for interesting central extension and Kac Moody algebra that may emerge with suitable choice of boundary conditions, in non-Abelian gauge theories for example \cite{Strominger1,Mitra}.  
Our method is also particularly suited for exploring subleading/subsubleading soft theorems (for e.g.\cite{Lysov2, Campiglia3, Campiglia4}) in a controlled manner. We hope to return to these questions in a future publication.

\section*{Acknowledgements}

Authors thank David Skinner, Malcolm Perry, Jorge Santos, Nick Dorey, Arif Mohd and Chen-Te Ma for valuable suggestions and correspondence. AB thanks Srijit Bhattacharjee  for  discussions on asymptotic symmetries and collaborations on related topics.  AB and LYH acknowledge support from Fudan University and Thousand Young Talents Program. Y. J. was supported by FDUROP, National University Student Innovation Program, and Hui-Chun Chin and Tsung-Dao Lee Chinese Undergraduate Research Endowment (Grant No. 15046).

\appendix

\section{Schwinger quantization procedure for Maxwell theory on a spacelike hypersurface}
We briefly review  Schwinger quantization procedure following  \cite{Frolov}  for the  Maxwell theory on a spacelike hypersurface. 
We take the Minkowski metric as
\be \label{eq104}
ds^2=-dt^2+dx^2+dy^2+dz^2,
\ee 
and take a constant t slice as our spacelike hypersurface, then in order to use the Schwinger quantization procedure, we rewrite the usual maxwell action 
\be
S=-\frac{1}{4}\int d^4 x \sqrt{-g} \mathcal{F}_{\mu\nu}\mathcal{F}^{\mu\nu}
\ee
 in a first order form
  \be \label{maxwellS}
  S[\mathcal{A}_{\mu},\mathcal{F}_{\mu\nu}]=\frac{1}{4}\int \Big[\mathcal{F}_{\mu\nu}\mathcal{F}^{\mu\nu}-\mathcal{F}^{\mu\nu}(\partial_{\mu}\mathcal{A}_{\nu}-\partial_{\nu}\mathcal{A}_{\mu})+\frac{1}{\sqrt{-g}}\partial_{\mu}(\sqrt{-g}(\mathcal{F}^{\mu\nu}-\mathcal{F}^{\mu\nu}))\mathcal{A}_{\nu}\Big]\sqrt{-g} d^4 x.
  \ee
where $\mathcal{F}_{\mu\nu}$ and $\mathcal{A}_{\mu}$ are treated as independent fields.
And then we vary our action with the boundaries of the two fields fixed under variation, then we get two equations of motion
  \be
  \mathcal{F}_{\mu\nu}=\partial_{\mu}\mathcal{A}_{\nu}-\partial_{\mu}\mathcal{A}_{\nu}.
  \ee
  \be
  \frac{1}{\sqrt{-g}}\partial_{\mu}(\sqrt{-g}\mathcal{F}^{\mu\nu})=0.
\ee  
The next step is to take the equation of motions while relaxing the variation of the fields on the boundary surfaces $\Sigma_0$ and $\Sigma_1$, then we will get
  \be
  \delta S(\mathcal{A}_{\mu},\mathcal{F}_{\mu\nu})=G_{\Sigma_1}-G_{\Sigma_{0}}.
  \ee
where the generator $G_{\Sigma}$ is $ 
  G_{\Sigma}=\frac{1}{2}\int_{\Sigma}(\mathcal{A}_{\nu}\delta \mathcal{F}^{\mu\nu}-\mathcal{F}^{\mu\nu}\delta A_{\nu})d\Sigma_{\mu}.$
The essence of Schwinger quantization procedure is that then we will get commutators for any independent field operators by requiring $[\mathcal O,G_{\Sigma}]=\frac{i}{2}\delta \mathcal{O}$. Here, we do it for the constant t slice by imposing temporal gauge
\be
\mathcal{A}^{0}=0
\ee
then for the independent fields, imposing
  \be
  [\mathcal{A}_{\mu},G_{\Sigma}]=\frac{i}{2}\delta \mathcal{A}_{\mu}, [\mathcal{F}^{\mu\nu}d\Sigma_{\mu},G_{\Sigma}]=\frac{i}{2}\delta \mathcal{F}^{\mu\nu}d\Sigma_{\mu}.
  \ee
means that 
\be
[A_i(\vec{x}), E_j(\vec{x'})]=-\delta_{i,j}\delta(\vec{x}-\vec{x'})
\ee
where $E_j=-\partial_0 A_j$ in this gauge, and this reproduces the usual commutation relation. Further subtleties arises as usual for gauge theories from the residue gauge degree of freedom and the properties of this constraint system. For example, we can eliminate the extra degree of freedom by imposing second class constraints to modify the commutators or imposing first class constraints to constrain our Hilbert space, and these will not be discussed here, but can be found in detail in \cite{Henn}

\section{Schwinger brackets with the stronger boundary condition $F_{uz}\vert_{u\to \pm \infty}=F_{u\bar z}\vert_{u\to \pm \infty}=0$}

We sketch out briefly the details of the computation of the brackets for $F_{uz}=0$ condition.  For this case we have,
\begin{align}
\begin{split}
G_{\Sigma}=\frac{1}{2}\int dzd\bar z&\Big[-T\sum_{n\neq 0}(\bar \alpha_n (-1)^n)\delta d_0-T\sum_{n\neq 0}(\alpha_n (-1)^n)\delta \bar d_0\\&+\sum_{m\neq 0}\Big((-)^mT\bar d_0+\frac{i\, T^2}{m\,\pi}\bar \alpha_{-m}+\sum_{n\neq 0}\frac{i\, T^2}{2\,\pi}(-1)^{m+n}(\frac{1}{n}-\frac{1}{m})\bar\alpha_{n}\Big)\delta \alpha_{m}\\&+\sum_{m\neq 0}\Big((-)^mT d_0+\frac{i\, T^2}{m\,\pi} \alpha_{-m}+\sum_{n\neq 0}\frac{i\, T^2}{2\,\pi}(-1)^{m+n}(\frac{1}{n}-\frac{1}{m})\alpha_{n}\Big)\delta \bar \alpha_{m}\Big].
\end{split}
\end{align}
Now we have four independent fields $d_0, \bar d_0, \alpha_n,\bar \alpha_n.$ Then we demand,
\begin{align}
\begin{split}
[d_0,G_{\Sigma}]=\frac{i}{2}\delta d_0,[\bar d_0,G_{\Sigma}]=\frac{i}{2}\delta \bar d_0,[\alpha_n,G_{\Sigma}]=\frac{i}{2}\delta \alpha_n,[\bar \alpha_n,G_{\Sigma}]=\frac{i}{2}\delta \bar \alpha_n.
\end{split}
\end{align}
This gives us the following solutions for the brackets,
 \begin{align}
 \begin{split}
 &[\alpha_n,\bar d_0]=[\bar \alpha_n,d_0]=-\frac{i}{T}(-1)^n,\\&
 [\alpha_n,\bar \alpha_m]=\frac{n \pi}{T^2}\delta_{m-n}+\frac{\pi}{3 T^2} n (-1)^{m+n}+\frac{\pi}{T^2}m (-1)^{m+n}.
 \end{split}
 \end{align}
Here we have used the fact that $\zeta(0)=-\frac{1}{2},$ $\zeta(s)=\sum_{n=1}^{\infty}\frac{1}{n^s}$ is the Riemann zeta function.

\end{document}